\documentclass[twocolumn,floatfix,aps,superscriptaddress,nofootinbib]{revtex4-1}
\usepackage{graphicx}
\usepackage{amsmath}
\usepackage{amssymb}
\usepackage{bm}

\begin{document}
\title{Gate-controlled Spin Extraction from Topological Insulator Surfaces}

\author{Ali Asgharpour}
\affiliation{Faculty of Engineering and Natural Sciences,
Sabanc{\i} University, Orhanl{\i} - Tuzla, 34956, Turkey}

\author{Cosimo Gorini}
\affiliation{Institut f\"ur Theoretische Physik, Universit\"at Regensburg, 93040 Regensburg, Germany}

\author{Sven Essert}
\affiliation{Institut f\"ur Theoretische Physik, Universit\"at Regensburg, 93040 Regensburg, Germany}

\author{Klaus Richter}
\affiliation{Institut f\"ur Theoretische Physik, Universit\"at Regensburg, 93040 Regensburg, Germany}

\author{\.{I}nan\c{c} Adagideli}
\email{adagideli@sabanciuniv.edu}
\affiliation{Faculty of Engineering and Natural Sciences,
Sabanc{\i} University, Orhanl{\i} - Tuzla, 34956, Turkey}

\date{\today}

\begin{abstract}

Spin-momentum locking, a key property of the surface states of
three-dimensional topological insulators (3DTIs), provides a new avenue
for spintronics applications. One consequence of spin-momentum locking is the induction of surface spin accumulations due to applied electric
fields. In this work, we investigate the extraction of such electrically-induced spins from their host TI material into  adjoining conventional, hence topologically trivial, materials that are commonly used in electronics devices. We focus on effective Hamiltonians for
bismuth-based 3DTI  materials in the ${\rm Bi}_2{\rm Se}_3$ family, and
numerically explore the geometries for extracting current-induced spins
from a TI surface. In particular, we consider a device geometry in which a side pocket is attached to various faces of a 3DTI quantum wire and show that it is possible to create current-induced spin accumulations in these topologically trivial side pockets. We further study how such spin extraction depends on geometry and material parameters, and find that electron-hole degrees of freedom can be utilized to control the polarization of the extracted spins by an
applied gate voltage.

\end{abstract}

\pacs{Valid PACS appear here}
\maketitle

\section{Introduction}
\label{sec_intro}

The push towards the utilization of the electron's spin degree of
freedom in common electronic devices,  which are conventionally based
on the manipulation of the electron charge,
has matured to the field called
spintronics~\cite{vzutic2004spintronics}. The various lines of research
in this field not only comprise questions of fundamental interest in
spin physics, but also focus on applications. Possible advantages of
utilizing spin-based elements in comparison to charge-based electronic
devices might be low power consumption and less heat dissipation, as
well as more compact and faster reading/writing of data.

The ferromagnets~\cite{ralph2008spin,
wolf2001spintronics,sato2002first} are the mainstream materials used in
spintronics where the ferromagnetic exchange interaction causes the
spin-dependency of transport, allowing the
creation/manipulation/detection of spins. However,  after the
celebrated Datta-Das spin transistor
proposal~\cite{datta1990electronic}, it became clear that spin-orbit
interaction can also be utilized for spin manipulation in electronic
devices. As the Datta-Das setting still requires ferromagnetic leads, a
parallel approach utilizing materials without intrinsic magnetism, such
as paramagnetic metals and semiconductors with only spin-orbit
coupling~\cite{awschalom2013semiconductor,
kato2004observation,Schliemann2017}, has become an attractive
alternative.

Various methods of spintronics implementations without ferromagnets
have emerged and developed over the recent
years~\cite{edelstein1990spin,inoue2003diffuse,
silov2004current,ganichev2004can, ganichev2006electric,
d1971possibility, dyakonov1971current, hirsch1999spin,
governale2003pumping, mal2003spin, murakami20042,Scheid-et-al}. These
methods are commonly based on (i)~the spin Hall
effect~\cite{dyakonov1971current}, where an applied electric current
generates a transverse spin current, and (ii)~ Edelstein (or inverse spin galvanic) effect~\cite{edelstein1990spin,aronov1989}, where an
applied electrical current generates a nonzero spin accumulation. Once
generated, as these spins drive spintronics circuits, they need to be
further manipulated and ultimately detected. For detection, inverse
effects corresponding to those mentioned above, namely the inverse spin
Hall effect~\cite{saitoh2006conversion, kimura2007room,
uchida2008observation, valenzuela2006direct, seki2008giant} and spin
galvanic effect (SGE)~\cite{ganichev2002, sanchez2013spin,
adagideli2007extracting, vzutic2002spin, shen2014microscopic} have been
successfully utilized. Main methods for spin manipulation are based on
exchange and Zeeman fields or spin-orbit coupling to induce spin
precession. However, weak coupling requires long length scales over
which the induced spins need to remain coherent. This is an issue as
spin precession lengths are usually comparable to spin
relaxation/dephasing lengths. Furthermore, the spin-orbit coupling
needs to be controlled over the precession (hence manipulation) region,
while spin-generation in part of the circuit needs to remain
unaffected. Hence in order to close the creation/manipulation/detection
cycle reliably, additional electrical methods for spin manipulation is
desirable.

\begin{figure}[h!]
	\centerline{\includegraphics[width=0.9\linewidth]{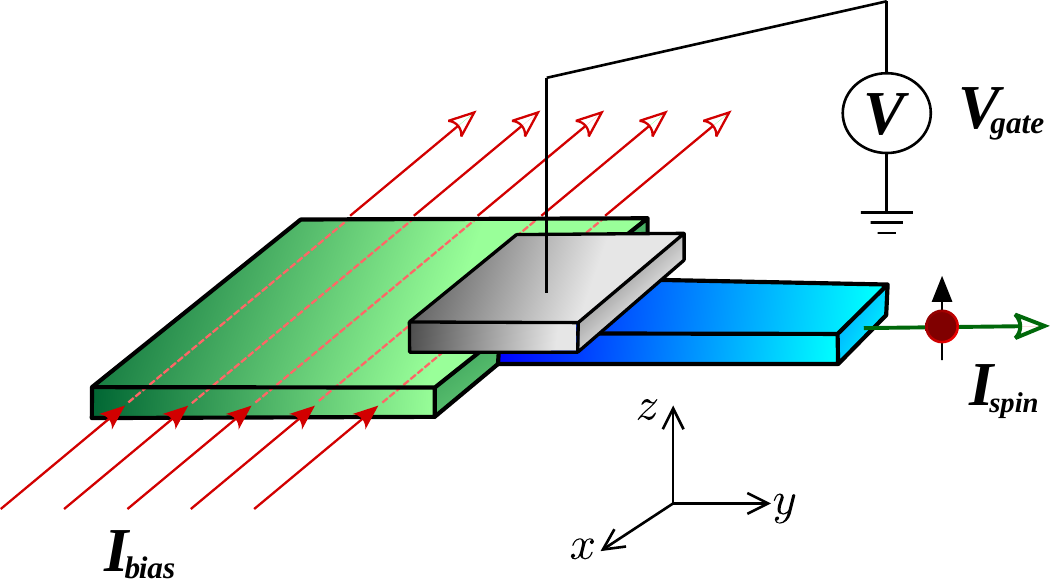}}
	\caption{(Color online) Slab of a topological insulator (green), current biased with $I_{\rm bias}$. The induced
		spin accumulation at the boundaries can be injected into a side contact (blue). A gate potential
		$V_{\rm gate}$ can be tuned to control the spin polarization of the spin injected current.}
	\label{fig:fig_layout}
\end{figure}

In this work, we consider a mechanism in TIs
that allows for local and all-electrical control of electrically
generated spins with gates. In most spintronics (or spin-orbitronics)
platforms, charge carriers are of a given type: either electron or
hole, implying that local application of gates equally couples to both
spin species. In others where electron and hole pockets might co-exist,
there is no coherence between the electron/hole degree of freedom and
the spin degree of freedom. As a consequence, electric gates cannot
locally control local spin accumulations in conventional spintronics
and spin-orbtronics platforms. On the other hand, the surface (or edge)
of 3D (2D) TIs feature both electron- and hole degrees of freedom as
well as spin-orbit coupling. Applied gates control the local potential,
which couples oppositely to electrons and holes, and spin-orbit coupling
allows for spin dependency of electron-hole degrees of freedom. We
demonstrate below that this joint property allows for electronic
control of spins locally {\em within a region much smaller than the
spin precession length}, the lengthscale over which spins can be manipulated in conventional spintronics applications~\cite{vzutic2004spintronics}.

As an explicit example, we consider 3DTI materials of the ${\rm
Bi}_2{\rm Se}_3$ family whose effective model is extensively discussed
in the literature \cite{Zhang, zhang2012, silvestrov2012, brey2014}.
Qualitatively, our conclusions should apply also to strained (3D)
\rm{HgTe}, though an equally successful effective model for such a
system is still missing. We focus on a particular geometry (sketched in
Fig.~\ref{fig:fig_layout}) and demonstrate how the spin extraction can
be controlled in a region smaller than the spin precession length. In
this geometry, the spins are generated by the spin galvanic effect at
the surface of the TI. By attaching a side pocket and tuning the
chemical potential on the pocket by an applied gate voltage, we
demonstrate that the extracted spins can change their polarization,
regardless of the generated spins on the TI side.

Our paper is organized as follows: In Sec.~\ref{subsec_stage}, we
outline the effective surface Hamiltonian of a 3DTI and the
corresponding spin operators. We then present the inverse spin galvanic
effect (ISGE), also known as Edelstein effect, through Kubo formalism in
Sec.~\ref{subsec_sgalvanic_basics}. Different names addressing the same phenomenon are used in the literature depending on context. In Sec.~\ref{subsec_paradox_solution}, we state an ISGE
paradox with its solution for the surfaces of a 3DTI. Next, we discuss
the model and the method proposed for extracting spin from surfaces of
a 3DTI in Sec.~\ref{subsec_model}. In Sec.~\ref{subsec_eigenmodes}, we
derive the spin behavior on the 3DTI surfaces, which we show to be in
close agreement with our numerical simulations. In
Sec.~\ref{subsec_extraction}, we demonstrate how to extract spins from
3DTI surfaces and how to manipulate their polarization through a gate
potential. We close with concluding remarks in
Sec.~\ref{sec_conclusions}.

\section{A spin-galvanic paradox and its solution}
\label{sec_paradox}

\subsection{Setting the stage}
\label{subsec_stage}

Consider a finite crystal of an anisotropic 3DTI material, such as
${\rm Bi}_2{\rm Se}_3$, which in its TI phase hosts topologically protected metallic surface states. The
existence of these states, described by a single Dirac cone, were
confirmed experimentally by ARPES~\cite{chen2009experimental,
xia2009observation} and STS~\cite{alpichshev2010stm,
zhang2009experimental, cheng2010landau, hanaguri2010momentum,
alpichshev2011stm} measurements. Further experiments confirmed the
helical nature of such surface states~\cite{hsieh2009observation}. The anisotropy of these materials implies that the topological metallic
states existing on the different crystal faces will be described by
Dirac-like effective Hamiltonians featuring different spin
structures~\cite{zhang2012, silvestrov2012, brey2014}.  We are
interested in the consequences of the anisotropy of these materials on
the ISGE~\cite{aronov1989, ivchenko1990current, edelstein1990spin}, for recent discussions see~\cite{ganichev2016, gorini2017, ando2017}.

The states of the 2D helical surfaces of ${\rm Bi}_2{\rm Se}_3$ are
admixtures of electron- and hole-like states of different parity
($\pm$) and spin ($\uparrow\downarrow$), coming from Bi and Se
$p_z$-orbitals, $|P1^+_z, \uparrow\downarrow\rangle$ and $|P2^-_z,
\uparrow\downarrow\rangle$, respectively~\cite{Zhang}. As a
consequence, the real spin content of such states does not necessarily
coincide with the pseudospin degrees of freedom used to label them.
Hence, $\sigma_i$ ($i=x, y, z$) denote the Pauli operators
corresponding to the two bands at the surface (the pseudospin), while
$s_i$ are the spin operators within this restricted Hillbert space. The
most commonly ``known'' low-energy effective Hamiltonian
for the topological surface state is that of the ``top'' and ``bottom''
surfaces in the growth direction, which we choose to be in the
$\hat{\bf z}$ direction:
\begin{equation}
H^{\pm\hat{\bf z}}=E_0({\hat{\bf z}}) + v_F({\hat{\bf z}}) \big({\bf k}\times\hat{\bf z}\big)\cdot{\boldsymbol \sigma},
\label{h1}
\end{equation}
where $E_0({\hat{\bf z}})$ is the energy of the Dirac point, $v_F({\hat{\bf z}})$ is the corresponding Fermi velocity and $\pm$ refers to the surface normals pointing away the bulk.
In this case the spin and the pseudospin operators are the same:
\begin{equation}
{\bf s}={\boldsymbol \sigma}.
\label{spinzsurface}
\end{equation}
This identification as well as the rotational symmetry, however is lost at the side surfaces:
\begin{equation}
H^{\pm\hat{\bf y}}=E_0({\hat{\bf y}}) \pm v_{F,x}({\hat{\bf y}}) k_x\sigma_z \mp v_{F,z}({\hat{\bf y}}) k_z\sigma_x,
\label{h2}
\end{equation}
where  $E_0({\hat{\bf y}})$ is the energy of the Dirac point, and $v_{F,x}({\hat{\bf y}})$ and $v_{F,z}({\hat{\bf y}})$ are the
corresponding Fermi velocity in the $x$ and $z$ directions, respectively. In this case, while the $x$ component of
the spin and the pseusospin operators are the same, they are merely proportional in the $\hat{\bf y}$ and $\hat{\bf z}$ surfaces with
the proportionality parameter $\eta$:
\begin{equation}
s_x = \sigma_x,\quad s_y=\eta \sigma_y,\quad s_z =\eta \sigma_z.  \label{spinysurface}
\end{equation}
For completeness, we express the $\pm \hat{\bf x}$ surface Hamiltonian as
\begin{equation}
H^{\pm\hat{\bf x}}=E_0({\hat{\bf x}}) \mp v_{F,y}({\hat{\bf x}}) k_y\sigma_z \pm v_{F,z}({\hat{\bf x}}) k_z\sigma_y,
\label{h3}
\end{equation}
\begin{equation}
s_x = \eta \sigma_x,\quad s_y= \sigma_y,\quad s_z =\eta \sigma_z, \label{spinxsurface}
\end{equation}
where $E_0({\hat{\bf x}}) = E_0({\hat{\bf y}})$, $v_{F,y}({\hat{\bf x}})$ and $v_{F,z}({\hat{\bf x}})$ are the Fermi velocities in the $y$ and $z$ directions, respectively.
To summarize, the real spin coincides with the Pauli matrices $\sigma_i, i=x,y,z$ of the pseudospin only on the $\pm\hat{\bf z}$ surface.
In particular, if $\eta\rightarrow0$, the surface states on the $\pm\hat{\bf y}$ side have $s_y=0, s_z=0$.
This point is crucial, as we discuss below.

\subsection{Spin galvanic basics}
\label{subsec_sgalvanic_basics}

We consider the spin accumulation, $s_z(\omega)$, generated in response to an applied
electric field $E_x$ in a spin-orbit coupled 2D system lying in the
$\hat{\bf x}$-$\hat{\bf z}$ plane -- corresponding to the side surfaces
$\pm\hat{\bf y}$. The ISGE can be written in Kubo
form~\cite{inoue2003diffuse} as
\begin{eqnarray}
s_z(\omega) &=& \sigma_{\text{ISGE}}(\omega) E_x(\omega)
\\
&=& \langle\langle s_z ; J_x \rangle\rangle A_x(\omega),
\end{eqnarray}
where $\langle\langle s_z ; J_x \rangle\rangle = \frac{-i}{\hbar} \int_0^t\,\langle [s_z(t),J_x(0)] \rangle e^{i\omega t}{\rm d}t$
is the Kubo linear response kernel, $A$ is the vector potential and $\sigma_{\text{ISGE}}$ is the frequency-dependent ISGE conductivity. Thus
\begin{equation}
\sigma_{\text{ISGE}}(\omega) = \frac{\langle\langle s_z ; J_x \rangle\rangle}{i\omega}.
\end{equation}
Its Onsager reciprocal effect, the spin galvanic effect (SGE), reads \cite{shen2014microscopic}
\begin{eqnarray}
J_x(\omega) &=& \sigma_{\text{SGE}}(\omega) {\dot B}_z(\omega)
\\
&=& \langle\langle J_x ; s_z \rangle\rangle B_z(\omega),
\label{eq:B}
\end{eqnarray}
yielding
\begin{equation}
\sigma_{\text{SGE}}(\omega) = \frac{\langle\langle J_x ; s_z \rangle\rangle}{i\omega}.
\end{equation}
In Eq.~(\ref{eq:B}) ${\dot B}$ is the time derivative of the magnetic field
which generates the non-equilibrium $s_z$ leading to the SGE.

\subsection{Spin galvanic effect on the surface of a 3DTI}
\label{subsec_paradox_solution}

As we stressed above, the relation between the pseudospin $\boldsymbol{\sigma}$ and the real
spin ${\bf s}$ on the 3DTI surface can be anisotropic.
The two quantities are identical on the $\pm\hat{\bf z}$ surfaces, and hence there is no ambiguity in calculating
the ISGE and the SGE on the surfaces. However, on the $\hat{\bf y}$ surfaces
\begin{equation}
\label{slambda1}
s_z = \eta \sigma_z.
\end{equation}

On the surface of the TI, spin and charge/momentum are locked. To be explicit we assume
\begin{equation}
\label{sclock1}
J_x = v_{F,x}({\hat{\bf y}}) \sigma_z
\end{equation}
with $v_{F,x}({\hat{\bf y}})$ is the Fermi velocity in the $x$-direction (see Eqs.~\eqref{h1}-\eqref{h2}).
From Eqs.~\eqref{slambda1} and \eqref{sclock1} one gets
\begin{equation}
\label{sclock2}
J_x = \frac{v_{F,x}({\hat{\bf y}})}{\eta} s_z.
\end{equation}
Equation~\eqref{sclock2} seems to imply a divergent (``colossal'') SGE for $\eta\rightarrow0$,
while the ISGE should vanish.

This apparent paradox is resolved by judiciously inspecting the SGE and ISGE linear response kernels.
First, for the SGE one has
\begin{eqnarray}
J_x &=& \frac{\langle\langle J_x ; s_z \rangle\rangle}{i\omega} B_z(\omega)
\\
&=& \eta v_{F,x}({\hat{\bf y}}) \underbrace{\frac{\langle\langle \sigma_z ; \sigma_z \rangle\rangle}{i\omega}}_{L_{\sigma\sigma}} B_z(\omega),
\end{eqnarray}
which tends to zero for $\eta\rightarrow0$ as it should: The pseudospin-pseudospin response function $L_{\sigma\sigma}$
defined above has no divergencies.
Similarly for the ISGE holds
\begin{eqnarray}
s_z &=& \frac{\langle\langle s_z ; J_x \rangle\rangle}{i\omega} E_x(\omega)
\\
&=& \eta v_{F,x}({\hat{\bf y}}) \underbrace{\frac{\langle\langle \sigma_z ; \sigma_z \rangle\rangle}{i\omega}}_{L_{\sigma\sigma}} E_x(\omega)
\end{eqnarray}
which is given by the same response function $L_{\sigma\sigma}$ and again vanishes in the $\eta\rightarrow0$ limit.

\section{Spin extraction from 3DTI surfaces}
\label{sec_extraction}

Even though it turns out that there is no paradox in the form of a
divergent SGE response, there are interesting consequences when
considering $\eta\rightarrow0$. In particular, as we show below,  it is
possible to extract current-induced spins from the side surfaces even
if these are not spin polarized. The main idea is the following: at the
side surfaces of a TI, an analytical examination of the non-equilibrium
population of the $k_x$ states (induced by, say, an applied bias)
reveals their composition to be a mixture of spin-up electron-like
and a spin-down hole-like quasiparticles whose spins partially
cancel each other. This is the origin of the parameter $\eta \ne 1$  in
general. In the limit $D_2\rightarrow0$ (hence $\eta\rightarrow0$) the
cancellation is perfect. Therefore, it suffices to contact the surface
with a ``pocket'' containing electrons or holes--in practice, a gated
semiconductor--so that only the spin-polarized electron- or hole-like
part of the surface state will leak out of the TI. A side pocket/lead
thus acts as a gate-tunable spin extractor: The sign of the extracted
spins can be reversed by simply switching the pocket polarity from $n$-
to $p$-type or vice versa, allowing for local electrical control of
spin polarization. Note the crucial observation that the size of the
region where the spin is reversed can be shorter than the spin
precession length (see Fig.~\ref{fig:halfehalfh} below).

\subsection{Model and method}
\label{subsec_model}

In the rest of this Section, we further study the spin extraction
effect through analytical and numerical means for 3DTI nanowires. The wires are described
by a 3D effective Hamiltonian which captures the basic low-energy
properties of ${\rm Bi}_2{\rm Se}_3$ family, including e.g.~${\rm
Bi}_2{\rm Se}_3$, ${\rm Bi}_2{\rm Te}_3$ and ${\rm Sb}_2{\rm Te}_3$
materials~\cite{Zhang,liu2010model}:
\begin{eqnarray}\label{eq_3dhamiltonian}
H^{3D} &=& E(\mathbf{k}) (\sigma_{0} \tau_{0}) + \textit{M}(\textbf{k}) (\sigma_{0} \tau_{z}) + A_1 \sin k_z (\sigma_{z} \tau_{x}) \nonumber \\ &+& A_2 (\sin k_x (\sigma_{x} \tau_{x}) + \sin k_y (\sigma_{y} \tau_{x})),
\end{eqnarray}
where
\begin{align}
\textit{M}(\textbf{k}) &= M_{0} - 2B_{2}(2 - \cos k_x - \cos k_y) - 2B_{1} (1-\cos k_z), \nonumber \\
E(\textbf{k}) &= C + 2D_{2}(2 - \cos k_x - \cos k_y) + 2D_{1} (1-\cos k_z).\nonumber
\end{align}
Here, $\sigma_{x,y,z}$ and $\tau_{x,y,z}$ are the Pauli matrices, and
$\sigma_{0}$ and $\tau_{0}$ are the $2\times 2$ identity matrices in
spin and orbital space, respectively. If $(M_{0}/B_{1} > 0)$ then the
system is in the topologically nontrivial phase and Dirac-like surface
states form within the bulk band gap. For a wire, due to the size
quantization around the wire, the surface states form 1D channels and
the lowest 1D subband is gapped due to its non-trivial Berry phase
\cite{bardarson2010,ziegler2018}.

In order to find the current-induced spin polarization on the 3DTI
nanowire surfaces, we need the spin operators expressed in the basis
used to represent Eq.~\eqref{eq_3dhamiltonian}. The basis states are
hybridized states of the Se and Bi $p$ orbitals with even ($+$) and odd
($-$) parities, and spins up ($\uparrow$) and down ($\downarrow$),
namely $|P1_{z}^{+},\uparrow\rangle$, $-i|P2_{z}^{-},\uparrow\rangle$,
$|P1_{z}^{+},\downarrow\rangle$, $i|P2_{z}^{-},\downarrow\rangle$, in
that order. Then the spin operators in the basis of bulk states are
given by~\cite{brey2014}:
\begin{equation}
\label{spin_3D}
S_{x}= \sigma_{x} \tau_{z}, \quad S_{y}= \sigma_{y} \tau_{z}, \quad S_{z}= \sigma_{z} \tau_0.
\end{equation}
Using the explicit forms of the spin operators, Eqs.~\eqref{spin_3D}, we generalize the Kubo response kernel of effective
2D surface model of the previous section to the more realistic 3D
model~\eqref{eq_3dhamiltonian}:
\begin{eqnarray}
S_z(\omega) &=& \sigma_{\text{ISGE}}(\omega) E_y(\omega)
\\
&=& \langle\langle S_z ; J_y \rangle\rangle A_y(\omega)
\end{eqnarray}
with $S_z = \sigma_z \tau_0$.

The effective surface description is obtained by projecting in to the
space spanned by the surface modes. One thus obtains the effective
surface spin and Hamiltonian operators (see
Appendix~\ref{app_projection_spin}). These surface Hamiltonians and
modes for electrons on 3DTI faces defined by their normals $\pm\hat{\bf
x}, \pm\hat{\bf y}, \pm\hat{\bf z}$, were computed by Brey and Fertig
\cite{brey2014}. In our geometry, the relevant surfaces are
$\pm\hat{\bf z}$ and $\pm\hat{\bf y}$ where the projections of the spin
operators follow Eq.~\eqref{spinzsurface} and Eq.~\eqref{spinysurface},
yielding the effective Hamiltonians Eq.~\eqref{h1} and Eq.~\eqref{h2}, respectively.
The parameters of surface Hamiltonians are then obtained from
Eq.~\eqref{eq_3dhamiltonian}~\cite{brey2014} by projection. In
particular, the band crossing energies of the ${\hat{\bf z}}$ and
${\hat{\bf y}}$ surfaces (which are the relevant surfaces for our
choice of axes) are given by:
\begin{align}
E_0({\hat{\bf z}})& = C+\xi M_0, \\
E_0({\hat{\bf y}}) &= C+\eta M_0 ,
\end{align}
and the corresponding Fermi velocities are given by:
\begin{align} \label{Eq:Fermi_velocities}
v_F({\hat{\bf z}})& =  A_2\sqrt{1-\xi^2}, \\ \label{Eq:Fermi_velocities1}
v_{F,x}({\hat{\bf y}})&=A_2\sqrt{1-\eta^2}, \\ \label{Eq:Fermi_velocities2}
v_{F,z}({\hat{\bf y}})&=A_1 \sqrt{1-\eta^2},
\end{align}
where
\begin{equation}
\xi = D_1/B_1, \quad \eta = D_2/B_2.
\end{equation}

In our numerical study, we use the tight-binding representation of the
Hamiltonian in Eq.~\eqref{eq_3dhamiltonian} and focus on a a 3DTI wire
attached to two semi-infinite leads (see Fig.~\ref{fig:3DTI}(a)). We
evaluate nonequilibrium local spin densities $\langle S_i \rangle(m)
=\langle \psi_\alpha(m) | S_i | \psi_\alpha({m}) \rangle$ for each site
$m$, where $\psi_\alpha({m})$ is the wavefunction of the (occupied)
state $\alpha$ at site $m$ and $S_i$ are the spin operators defined in
Eq.~\eqref{spin_3D}. We then sum over all occupied states $\alpha$. For
an infinitesimal bias, these are all scattering wavefunctions at a
certain energy, $E_F$ originating from one of the leads, depending on
the sign of the bias. Local charge density is similarly obtained when
$S_i\rightarrow \sigma_{0} \tau_{0}$. We utilize the KWANT
toolbox~\cite{groth2014kwant} for our numerical simulations. The
parameters of our band Hamiltonian are chosen from ab-initio band
structure calculations of ${\rm Bi}_{2}{\rm
Se}_{3}$~\cite{liu2010model} in our numerical simulations. The
particular values used are $A_{1}=2.2 \, \rm{eV} \AA$, $A_{2}=4.1 \, \rm{eV}
\AA$, $B_{1}=10 \, \rm{eV} \AA^{2}$, $B_{2}=56.6\, \rm{eV} \AA^{2}$,
$C=-0.0068 \, \rm{eV}$, $D_{1}=1.3 \, \rm{eV} \AA^{2}$, $D_{2}=19.6 \,
\rm{eV} \AA^{2}$ and $M=0.28 \, \rm{eV}$. We have also set the lattice
constant to be $a = 5 \, \rm{\AA}$ in our numerical calculations.

\subsection{Spin dynamics and accumulation at the surface}
\label{subsec_eigenmodes}

\begin{figure}[h!]
	\centerline{\includegraphics[width=0.98\linewidth]{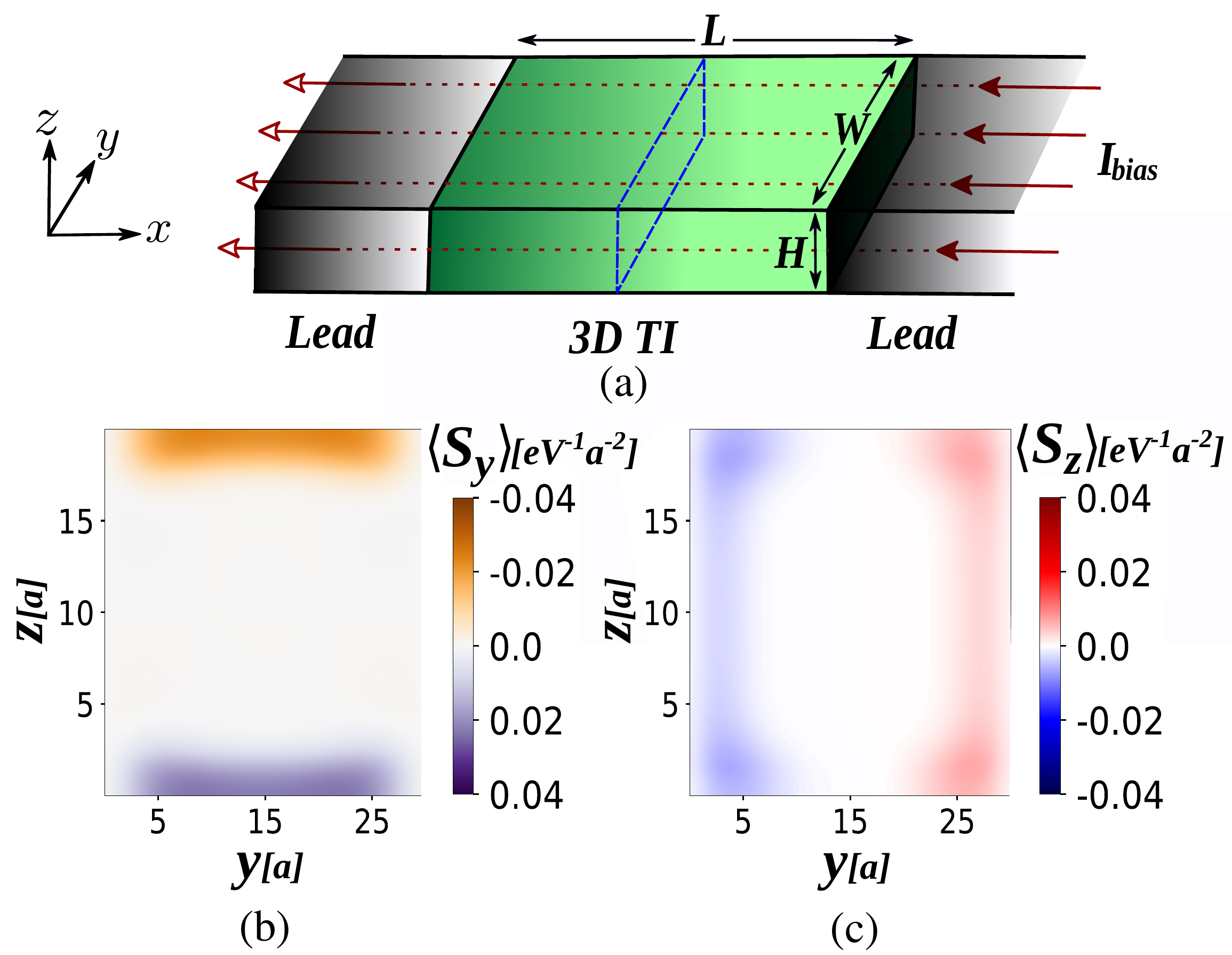}}
	\caption{\label{fig:3DTI}
		Surface spin polarization of a 3DTI nanowire.
		(a) Sketch of a 3DTI nanowire attached to two semi-infinite leads.
		(b) $\langle S_{y} \rangle$ and
		(c) $\langle S_{z} \rangle$ denote spatial profile of the averaged
		spin polarization (averaged over 1000 disorder configurations) along cross sections, oriented in $\hat{\bf x}$ direction and marked as the blue
		rectangle in panel a). Parameters used are: $L = 30 \, a$, $W = 30 \, a$, $H = 20 \, a$,
		$H_{SP}=10 \, a$, $U_{0}=0.5 \, \rm{eV}$ and $E_{F} = 0.15 \, \rm{eV}$ which is in the bulk gap.}
\end{figure}

As a consequence of the locking of the spin and the momentum of the surface states in 3DTIs, the dynamics of spin and charge distributions are coupled. Moreover, even nonmagnetic impurities can flip an electron's spin during scattering, leading to the dominant spin relaxation mechanism --a variant of the Dyakonov-Perel spin relaxation~\cite{dyakonov1972spin}. All these are summarized by the spin diffusion equations, valid at lengthscales much larger than the
mean free path, that describes the coupled dynamics of spin and charge.
For the top ($\hat{\bf z}$) surface of the TI, the relevant diffusion
equations are given by~\cite{burkov2010spin}:
\begin{eqnarray}
\label{burkov's main relations}
\dfrac{\partial \Sigma_{i}}{\partial t} + \dfrac{\Sigma_{i}}{\tau}&=& \dfrac{D}{2} \nabla^2 \Sigma_{i} + |\epsilon_{ij}| D \dfrac{\partial \,\Xi}{\partial x_j}
- \epsilon_{ij} \dfrac{v_F}{2} \frac{\partial n}{\partial x_j},  \nonumber \\
 \Xi &=&\dfrac{\partial \Sigma_x}{\partial y}+\dfrac{\partial \Sigma_y}{\partial x}, \\
\dfrac{\partial n}{\partial t} &=& D \nabla^2 n + v_F ({\hat{\bf z}} \times \bm{\nabla}) \cdot \mathbf{\Sigma}, \nonumber
\end{eqnarray}
where $\epsilon_{ij}$ is the totally antisymmetric tensor, $D =
\dfrac{v_F^2 \tau}{2}$ is the diffusion constant which is proportional
to mean free time, $\tau$, and Fermi velocity, $v_{F}$ (for $\hat{\bf  z}$ surface, $v_F = v_F (\hat{\bf  z})$). $\Sigma_i$ are the  components of the pseudospin nonequilibrium density, $\bf
\Sigma$,  $n$ is the charge density and $\pm$ refers to the top and
bottom surfaces. In order to apply Eq.~\eqref{burkov's main relations}
to the side surfaces, ($v_F = v_{F,x}(\hat{\bf y})$, see Appendix~\ref{app_mfp}), we generalize the diffusion equations to
anisotropic surfaces and obtain how the accumulated real spins depend
on the charge gradients due to applied voltage bias:
\begin{eqnarray}
\label{fraction}
\left( \frac{\langle S_z \rangle} {d \langle n \rangle /dx}\right)_{\pm \hat{\bf y}} &=& \mp \left( \eta\dfrac{v_{F,x}(\hat{\bf y}) \tau}{2}\right)_{\pm \hat{\bf y}}, \\
\label{fraction2}
\left( \frac{\langle S_y \rangle}  {d \langle n \rangle / dx} \right)_{\pm \hat{\bf z}} &=& \pm \left(\dfrac{v_F(\hat{\bf  z}) \tau}{2}\right)_{\pm \hat{\bf z}}.
\end{eqnarray}
Hence, if $E_{F}$ sits in the bulk gap, then applying a bias voltage
yields surface currents flowing in the $x$-direction, which in turn
induces spin accumulations on the $\pm\hat{\bf y}$ and the $\pm\hat{\bf
z}$ surfaces. This is the ISGE. In order to test these predictions, we
numerically obtain spin densities via the method described in
Sec.~\ref{subsec_model}. Our results are shown in
Figs.~\ref{fig:3DTI}(b) and ~\ref{fig:3DTI}(c), where we plot the
$x$-averaged cross-sectional profile for $\langle S_y\rangle$ and
$\langle S_z\rangle$. Note that both components of the spin
accumulation are localized to the respective surfaces and have opposite
sign on opposite surfaces. Notice also that $\langle S_x\rangle=0$ in
our configuration since it is along the current direction. Furthermore,
$\langle S_z\rangle$ is smaller than $\langle S_y \rangle$ for
$\eta < 1$. The case $D_{2}=0$, as mentioned earlier, corresponds
to a vanishing ISGE $\langle S_z\rangle$ and the ``paradoxical'' regime
$\eta=0$ of Sec.~\ref{sec_paradox}.

In order to test Eqs.~(\ref{fraction}) and (\ref{fraction2}) numerically, we consider the quotient on the left hand side of these equations
as a function of disorder strength $U_0$.
Since  in the golden rule regime $1/\tau \sim U_0^2$, we expect a $U_0^{-2}$ behavior. In order to get the exact relation, we analytically calculate
the mean free time using a
${\bf k}\cdot{\bf p}$ approximation for surface eigenmodes in Appendix~\ref{app_mfp}.
Next, we perform numerical simulations and obtain the local spin/charge accumulations and
avearge these over a square
region in the middle of the $+\hat{\bf z}$ and $-\hat{\bf y}$ surfaces as well as over different disorder configurations with strength $U_0$.
Finally, we compare our analytical prediction (the blue line) for
the left-hand-sides of Eqs.~(\ref{fraction}) and (\ref{fraction2}) against the numerical simulations (red dots)
in Figs.~\ref{fig:szoverdnpart} and \ref{fig:syoverdnpart}, respectively. We find that our numerical results for ISGE are well described by the
analytical formulas in Eqs.~(\ref{fraction}) and (\ref{fraction2}).

\begin{figure}[h!]
	\centerline{\includegraphics[width=0.98\linewidth]{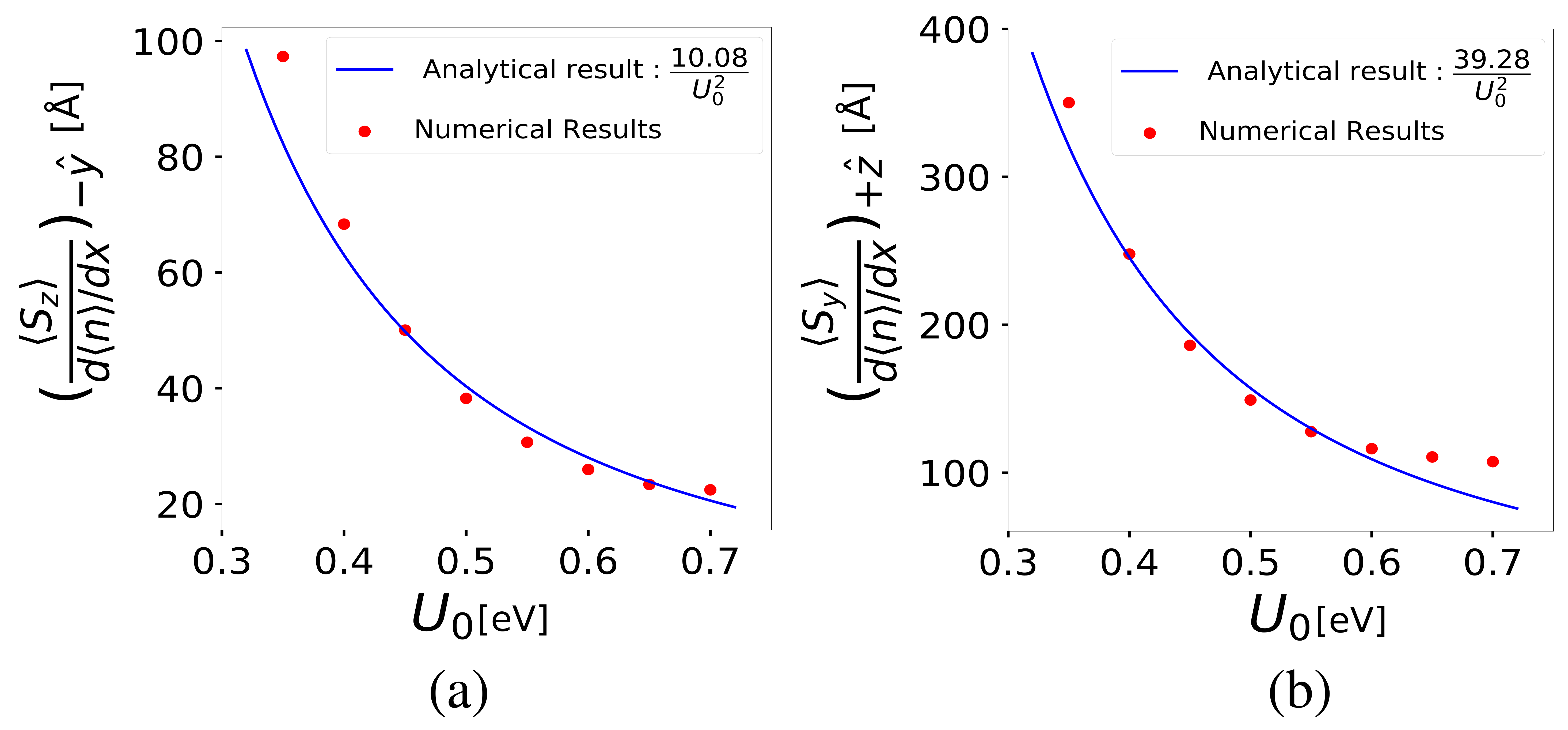}}
	\caption{Average ratios (a)
	$(\langle S_z \rangle / d \langle n \rangle / dx)_{-\hat{\bf y}}$ and
	(b) $(\langle S_y \rangle / d \langle n \rangle / dx)_{+\hat{\bf z}}$ as a function of disorder strength $U_0$. The blue curves show the analytical and the red symbols the numerical results. Parameters in our simulations are $L = 30 \, a$, $W = 30 \, a$, $H = 20 \, a$, and $E_{F} = 0.15 \, \rm{eV}$ which is in the bulk gap.}
\label{fig:szandsyoverdnpart}
\end{figure}

\subsection{Spin extraction}
\label{subsec_extraction}

\begin{figure}[h!]
	\centerline{\includegraphics[width=0.98\linewidth]{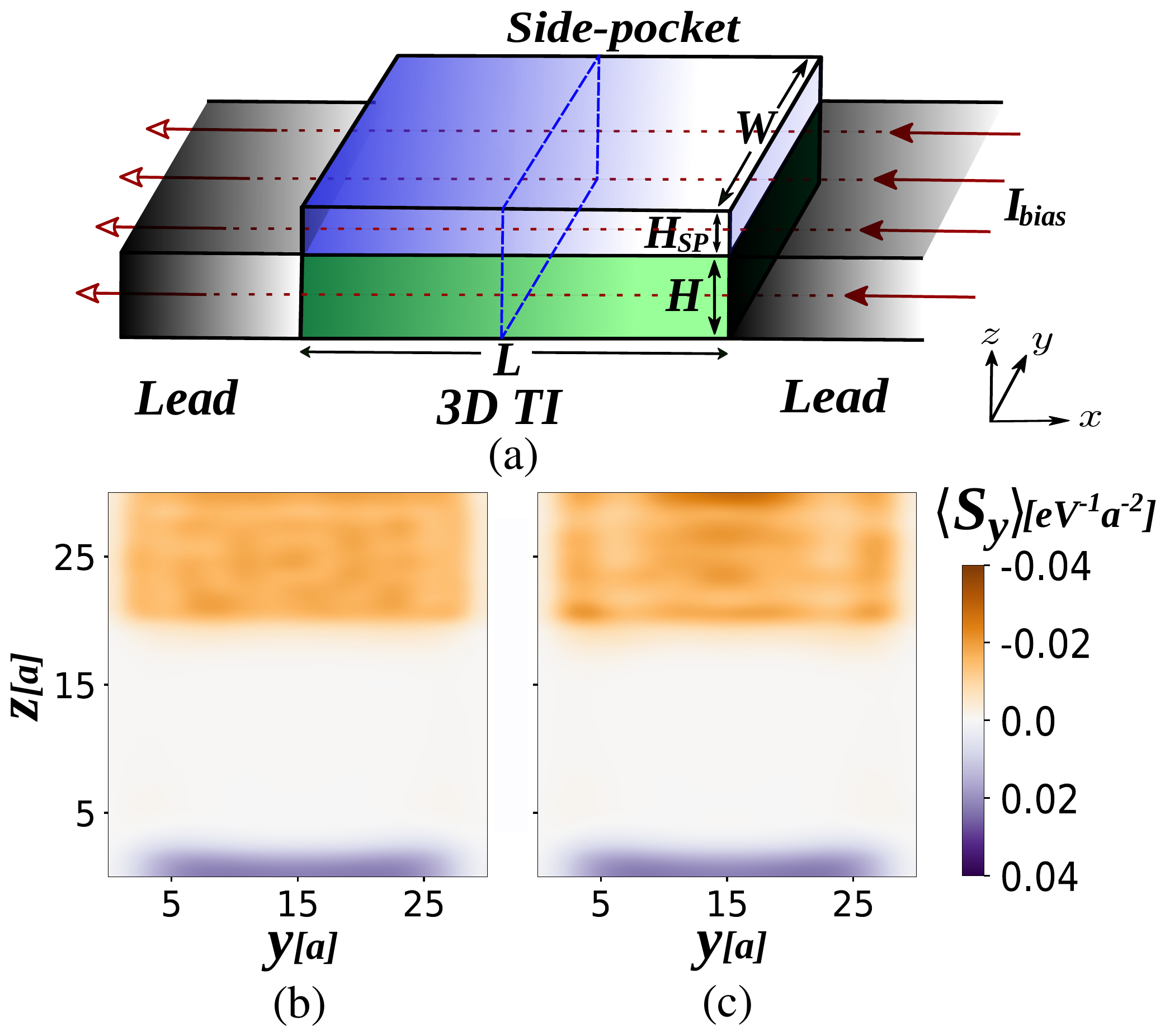}}
		\caption{\label{fig:topSP}
               Current-induced spin polarization into a side pocket at the top surface.
		(a)
		(b), (c) Spatial profile of the averaged spin polarization $\langle S_{y}(y,z) \rangle$ (averaged over 1000 disorder configurations)
		along cross sections in $\hat{\bf x}$ direction shown as dashed blue rectangle in panel a).
		In panels (b) and (c) the side pockets are doped to hole bands ($V_{\text{gate}}=-0.8 \, \rm{eV}$)
		and electron bands ($V_{\text{gate}}=0.9 \, eV$), respectively.
		Common parameters are $L = 30 \, a$, $W = 30 \, a$, $H = 20 \, a$, $H_{SP}=10 \, a$, $U_{0}=0.5 \, \rm{eV}$ and $E_{F} = 0.15 \, \rm{eV}$ which is in the bulk gap.}
\end{figure}

\begin{figure}[h!]
	\centerline{\includegraphics[width=0.98\linewidth]{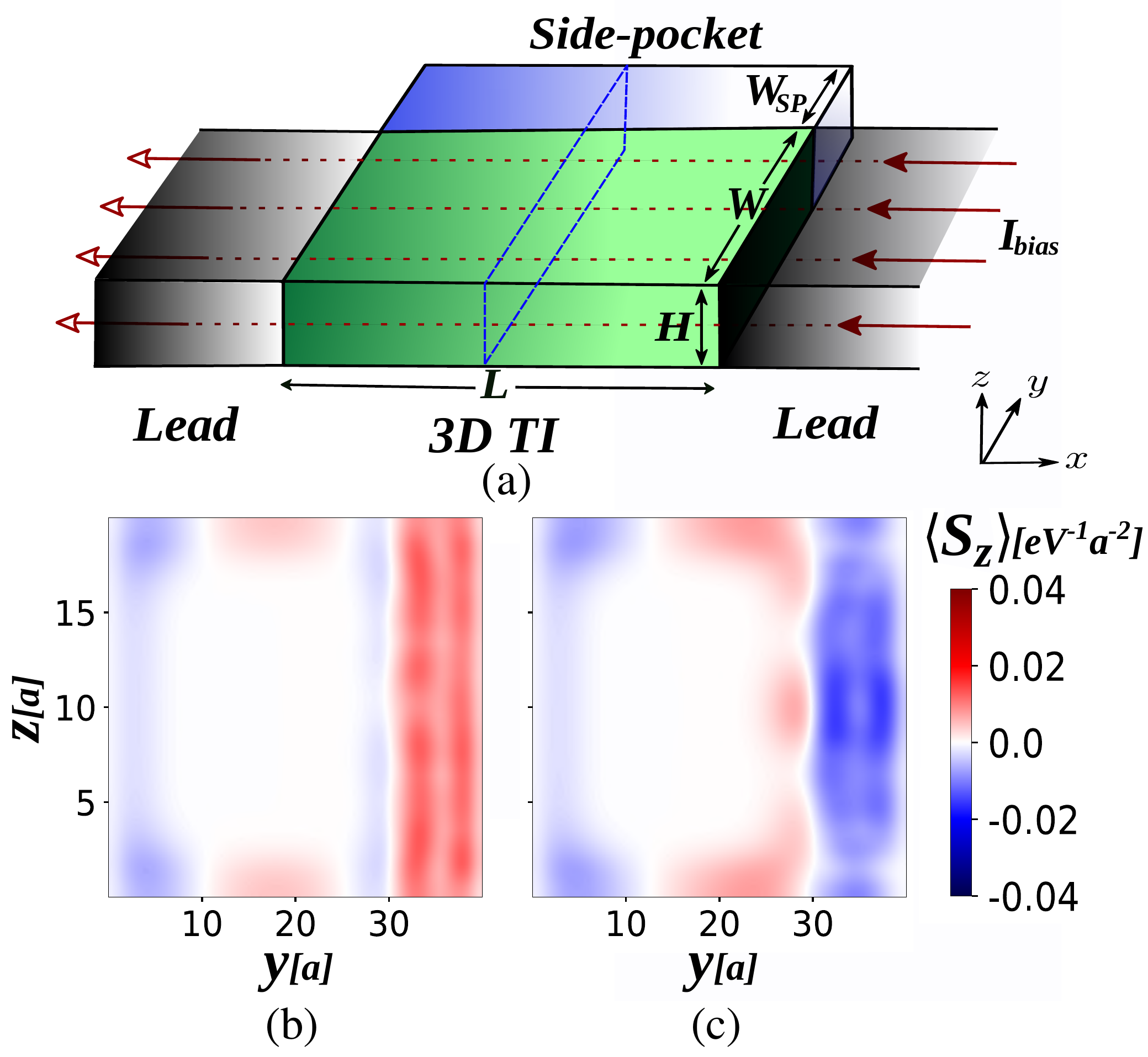}}
		\caption{\label{fig:sideSP}
               Current-induced spin polarization into a side pocket at the side surface.
		(a) Sketch of a side pocket attached to the side surface of the system shown in Fig.~\ref{fig:3DTI}(a).
		(b), (c) Spatial profile of the averaged spin polarization $\langle S_{z}(y,z) \rangle$ (averaged over 1000 disorder configurations)
		along cross sections in $\hat{\bf x}$ direction shown as dashed blue rectangle in panel (a).
		In panels (b) and (c) the side pockets are doped to hole bands ($V_{\text{gate}}=-0.8 \, \rm{eV}$)
		and electron bands ($V_{\text{gate}}=0.9 \, \rm{eV}$), respectively. Common parameters are: $L = 30 \, a$, $W = 30 \, a$, $H = 20 \, a$, $W_{SP}=10 \, a$, $U_{0}=0.5 \, \rm{eV}$ and $E_{F} = 0.15 \, \rm{eV}$ which is in the bulk gap.}
\end{figure}

Having discussed how spins can be induced at a topological insulator
surface, we now study how these spins can be extracted to be used in
(presumably topologically trivial) spintronics circuitry. To this end,
we focus on a geometry where a topologically trivial side pocket is
attached to the TI nanowire (see Figs.~\ref{fig:topSP}(a)
and~\ref{fig:sideSP}(a)). The current-induced spins at the TI surface
can then leak into the side pocket, generating nonzero spin
accumulation inside the side pocket. The nanowire size is chosen such
that its length and width $L=W=15{\rm nm}$ exceed the mean free path
$l$, ensuring diffusive carrier dynamics. The mean free path is
estimated in terms of the disorder potential strength $U_0$ using
Fermi's Golden Rule (see Appendix~\ref{app_mfp} for details). Note that
(pseudo)spin-charge locking implies that diffusion-like equations for
the spin can be employed, even though the spin dynamics is not
diffusive \cite{schwab2011}.

Spin extraction can take place at pockets that are attached to either surface of the 3DTI nanowire,
see Fig.~\ref{fig:topSP}(a) and Fig.~\ref{fig:sideSP}(a) for the geometry where the pocket is attached to the $\hat{\bf z}$ surface or the $\hat{\bf y}$ surface, respectively.
The pockets are gated in order to tune them to a metallic state, while charge carriers can be either electron- or hole-like states, thus coupling only to
the electron- or hole-like spin-momentum locked components of the 3DTI surface states. The gating is modeled by adding a corresponding on-site energy term in the tight-binding grid, while keeping the other
parameters of the effective Hamiltonian unchanged.

\begin{figure}[h!]
  \centerline{\includegraphics[width=0.95\linewidth]{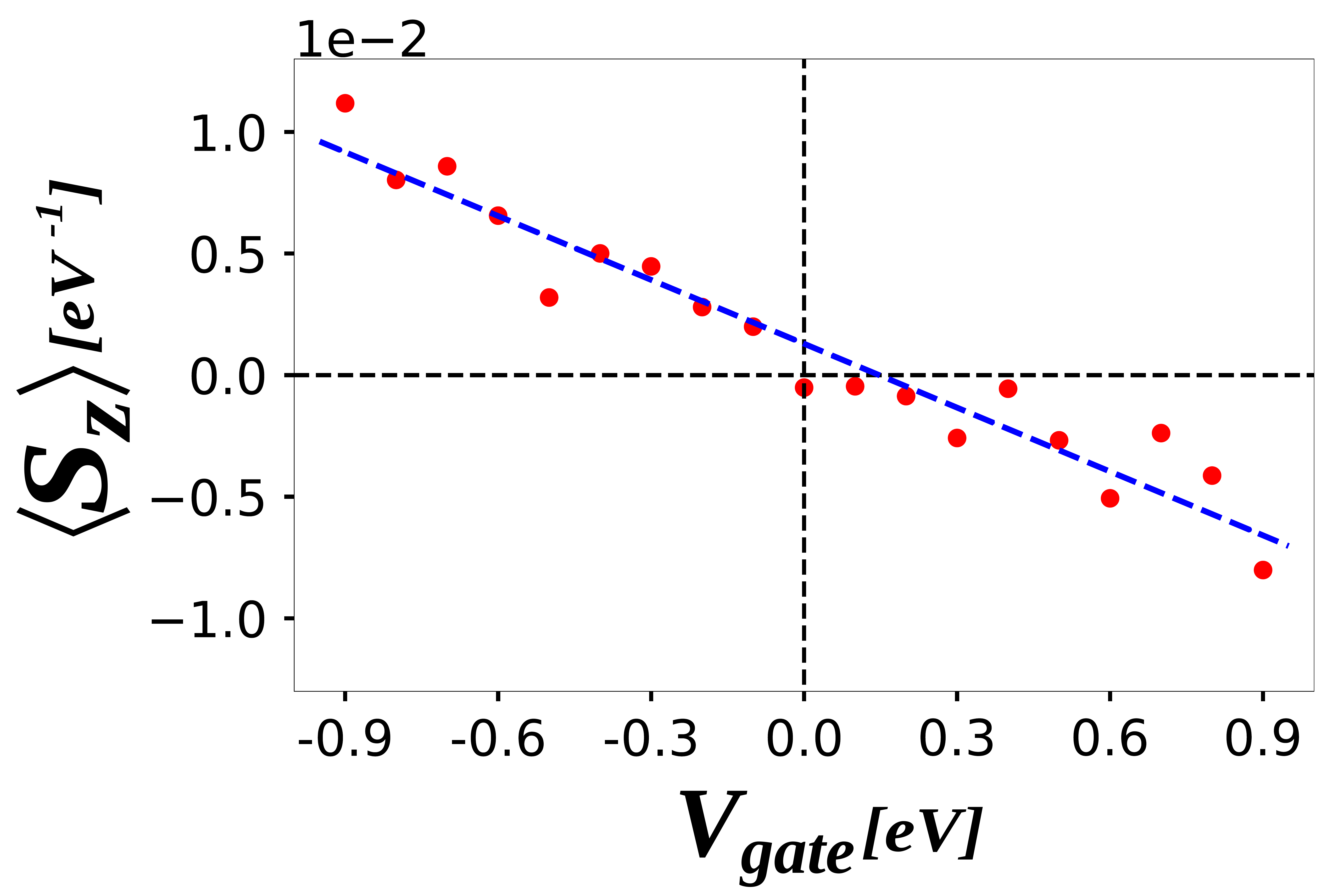}}
  \caption{
           Gate-dependence of side pocket spin polarization. $\langle S_{z} \rangle$ averaged over 1000 disordered configurations and doped side pocket sites is plotted versus gate potential. We consider the following parameters: $L=30 \, a$, $W=30 \, a$, $H=20 \, a$ $W_{SP}=10 \, a$, $U_{0}=0.5 \, \rm{eV}$ and $E_{F}=0.15 \, \rm{eV}$ which is in the bulk gap. The blue line is the best fitted line.}
  \label{fig:disorder}
\end{figure}

We perform tight-binding simulations and numerically calculate the current-induced spin polarization
$\langle S_{i} \rangle$, ($i=y,z$), averaging over 1000 disorder configurations
for a nanowire with side pockets.
Figs.~\ref{fig:topSP}(b),~\ref{fig:topSP}(c) and Figs.~\ref{fig:sideSP}(b),~\ref{fig:sideSP}(c)
show the spatial profile of the spin polarization along a perpendicular cross-section
for fixed doping values in hole and electron bands, respectively.
Focusing on the top ($\hat{\bf z}$) surface, our simulations show all expected features: A substantial non-equilibrium spin accumulation can
be extracted into the doped side pockets (Fig.~\ref{fig:topSP}). The extraction to the side ($\hat{\bf y}$) surface (Fig.~\ref{fig:sideSP}), on the other hand,
has non trivial features.
We first note the somewhat surprising fact that even if the 3DTI
surface has negligibly small spin accumulation, $\eta\approx 0$,
the spin accumulation extracted into the side pocket is nonnegligible (see corresponding figures in Appendix~\ref{app_zero_eta}).
Furthermore, the extracted spin polarization changes sign when the gate voltage is tuned so that the charge carriers change from
electrons to holes as can be seen from Figs.~\ref{fig:sideSP}(b) and \ref{fig:sideSP}(c).
We find that the geometry of the contact does not play a crucial role as it does for a 2D electron gas with Rashba spin-orbit coupling:
In that case wide contacts lead to reduced extraction~\cite{adagideli2007extracting} while for TIs wider contacts lead to enhanced extraction.
In order to further study the spin-gate effect mentioned above, we plot the spin accumulation $\langle S_z \rangle$ averaged over the side pocket,
as a function of the gate voltage applied to the side pocket, in Fig.~\ref{fig:disorder}. We find that the spin accumulation depends linearly on the gate voltage
and the sign of polarization changes by switching the side pocket polarity from hole- to electron-type.

\begin{figure}[h!]
	\centerline{\includegraphics[width=0.9\linewidth]{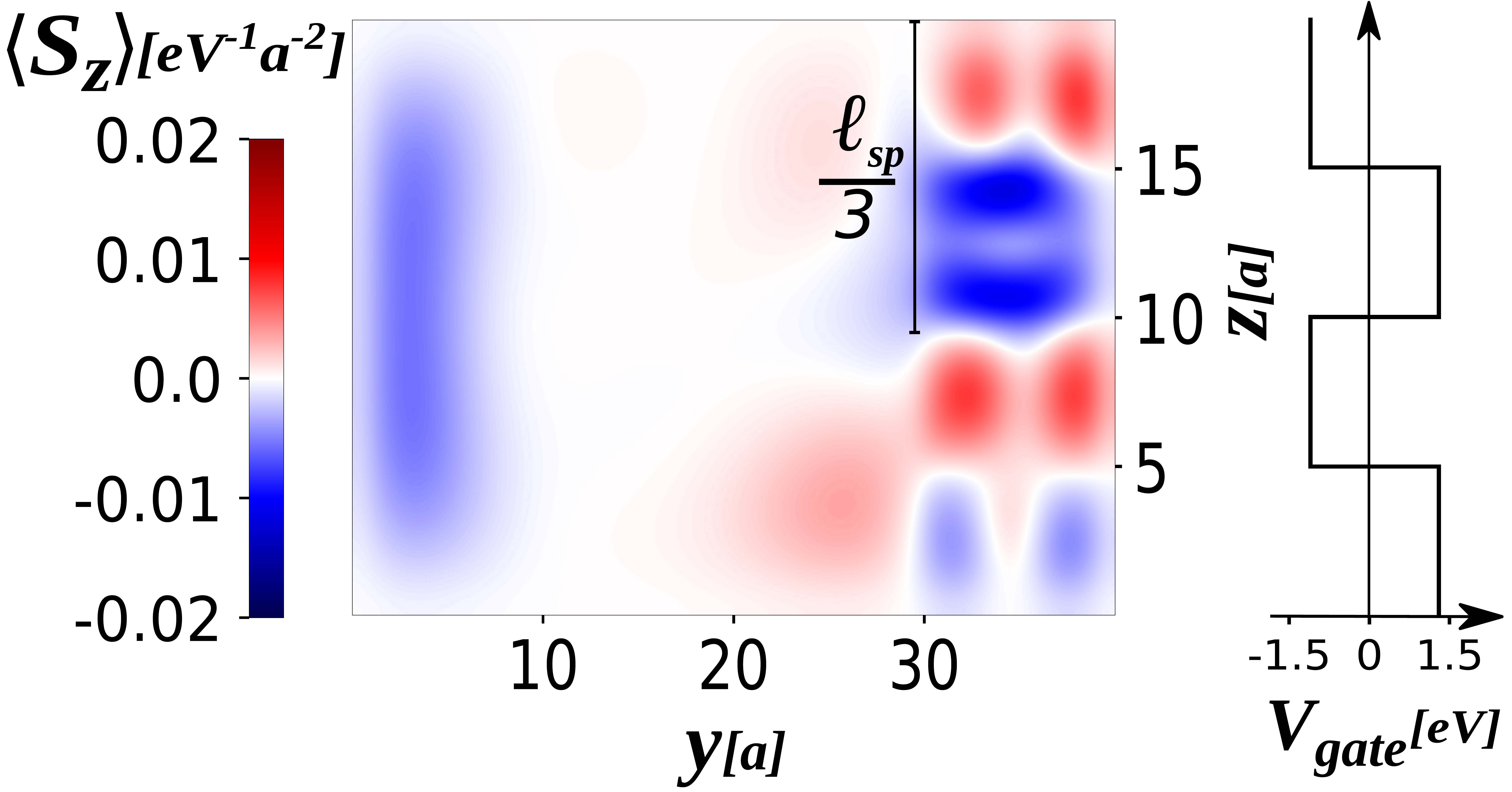}}
	\caption{Spatial profile of the averaged spin polarization $\langle S_{z}(y,z) \rangle$
		    (averaged over 1000 disorder configurations) along cross sections in $\hat{\bf x}$ direction for the system shown in Fig.~\ref{fig:sideSP}(a). Side pocket is alternatively doped to electron bands ($V_{\text{gate}}=1.3 \, \rm{eV}$) and hole bands ($V_{\text{gate}}=-1.1 \, \rm{eV}$). Side pocket is divided to four parts in $\hat{\bf z}$ direction and spacial profile is averaged over $\hat{\bf x}$ planes. Spin precession length in $\hat{\bf z}$ direction, $\ell_{sp} = 31.5 \, a$. Other parameters used are: $W_{SP}=10 \, a$, $L = 30 \, a$, $W = 30 \, a$, $H = 20 \, a$, $U_{0}=0.7 \, \rm{eV}$ and $E_{F} = 0.15 \, \rm{eV}$ which is in the bulk gap.}	
	\label{fig:halfehalfh}
\end{figure}

Finally we show that one can locally control the polarization direction of different parts of side pockets by local gating.
In Fig.~\ref{fig:halfehalfh}, we apply local gate profile where the electron puddles change into hole puddles within a region much smaller than the spin-precession length  $\ell_{sp}$. We find that the spatial profile of the polarization of the extracted spin accumulation, closely follow the local gate potential. Thus, we show that it is possible to electrically control local spin polarization within length scales much smaller than the spin precession length.

\section{Conclusions}
\label{sec_conclusions}

In conclusion, we focus on the current-induced spins at the surfaces of
3DTIs and show how to extract these spins into topologically trivial
materials commonly used in electronic devices. We find that unlike the
corresponding effect in 2D electron gases with Rashba spin-orbit
interaction, the mixing of the electron and hole degrees of freedom at
the TI surface allows for additional methods for spin manipulation. In
particular, we exposed a way to use electrical gate potentials to
locally manipulate spins in regions smaller than the spin precession
length. This opens up new possibilities for spin manipulation in
spintronics devices.

\section*{Acknowledgements}

A.A. thanks B. Pekerten and V. Sazgari for helpful conversations. This work was supported by Scientific and Technological Research Council of Turkey (TUBITAK) under Project Grant No.~114F163
and by the Deutsche Forschungsgemeinschaft (DFG, German Research Foundation) -- Project-ID 314695032 -- SFB 1277 (Subproject A07).


\appendix

\section{Effective surface Hamiltonians and spin operators}
\label{app_projection_spin}

Surface states in 3DTIs decay exponentially into the bulk and have
energies in the bulk bandgap. We first consider a semi-infinite 3DTI system
situated in $z \geq 0$ ($z \leq 0$) with a surface normal
$-\hat{\bf z}$ ($\hat{\bf z}$) pointing away from the bulk. By considering a
vanishing boundary condition at the surface, eigenfunctions
corresponding to these states can be written as
\begin{equation}\label{envelope}
\phi \sim u(k_{x}, k_{y}, \lambda_{1,2}) e^{i(k_{x}x + k_{y}y)} (e^{\pm \lambda_{1}z}-e^{\pm \lambda_{2}z}) \, ,
\end{equation}
where the $\pm$ sign in the $z$-direction corresponds to a system with a surface normal in the $\mp \hat{\bf z}$ direction at $z=0$. Here $Re (\lambda_{1,2}) > 0$ and $u(k_{x}, k_{y}, \lambda_{1,2})$ is a spinor that is an eigenstate of the 3DTI Hamiltonian described in Eq.~\eqref{eq_3dhamiltonian}, corresponding
to $k_{z}=-i\lambda_{1,2}$:
\begin{equation} \label{spinors}
\begin{aligned}
u^{\pm \hat{\bf z}}=&\dfrac{1}{\sqrt{2}}\begin{bmatrix}
\quad \sqrt{1 + \xi} \,\\
\mp i\sqrt{1 - \xi} \,\\
0 \\
0
\end{bmatrix}, \quad \quad
v^{\pm \hat{\bf z}} = \dfrac{1}{\sqrt{2}}\begin{bmatrix}
0 \\
0 \\
\quad \sqrt{1 + \xi} \,\\
\pm i \sqrt{1 - \xi} \,
\end{bmatrix}
\end{aligned}
\end{equation}
with energy dispersion to the lowest order of $k$ given by
\begin{equation} \label{energy}
E^\pm = C + \xi M_0 \pm A_2 \sqrt{1 - \xi^2} \, k_\perp,
\end{equation}
where $k_\perp^2 = k_x^2 + k_y^2$. Hence, the effective surface Hamiltonian as given in the text is obtained through projecting the 3DTI Hamiltonian in basis states given in Eq.~\eqref{envelope} and using the spinor eigenstates stated in Eq.~\eqref{spinors}. To lowest order in $k_x$ and $k_y$ , this results in
\begin{equation}
H^{\pm \hat{\bf z}} = C + \xi M_0 \pm A_2\sqrt{1\!-\!\xi^2}\,
\!\!
\left(
\!\!
\begin{array}{cc}
0 & ik_x + k_y \\
-ik_x + k_y & 0
\end{array}
\!\!\!
\right),
\end{equation}
which is introduced as Eq.~\eqref{h1} in the paper. The real spin operators for $\hat{z}$ surface are formed by projecting the spin operators in the basis of bulk states, Eq.~\eqref{spin_3D}, onto the two surface states
\begin{equation}
s_x = \sigma_x,\quad s_y=\sigma_y,\quad s_z = \sigma_z,
\end{equation}
which is stated as Eq.~\eqref{spinzsurface}. The effective surface Hamiltonians and real spin operators corresponding to other surfaces can be calculated similarly.

\section{Mean free time estimation}
\label{app_mfp}
We proceed with a Fermi's Golden Rule estimation of the mean free path.
The surface modes are 4-spinors with $k$-dependent components~\cite{shan2010} due to (pseudo)spin-momentum coupling.
Such $k$-dependence can lead to substantial differences between lifetime and transport time
\cite{culcer2010}. In the case of uncorrelated disorder, however, the difference is only an ${\mathcal{O}(1)}$-factor
\cite{schwab2011} and thus irrelevant for our estimations.
We thus work exclusively with band-bottom $k=0$ spinors.
We consider a TI slab extended in $x$ and $y$ directions, having a length $L$ and a width $W$ along $x$-direction and $y$-direction,
respectively, and a thickness $H$ along the $z$-direction. We further assume white-noise disorder of the form $\left\langle
V(\boldsymbol{r})\,V(\boldsymbol{r}')\right\rangle
=\gamma\,\delta(\boldsymbol{r}-\boldsymbol{r}')$. Therefore, using spinors stated in Eq.~\eqref{spinors} leads to
\begin{align}\label{vkk'}
\left \langle|V_{kk'}|^2\right \rangle= \dfrac{\gamma}{LW} \dfrac{\alpha}{\beta^2},
\end{align}
where $\alpha = \int_{0}^{H} {\rm d}z f^2(z)$, $\beta = \int_{0}^{H} {\rm d}z f(z)$ and $f(z)=(e^{-\lambda_1^* z} - e^{-\lambda_2^* z})(e^{-\lambda_1 z} - e^{-\lambda_2 z})$.
We use Fermi's Golden rule to derive the inverse mean free time and find
\begin{equation} \label{FGR}
\dfrac{1}{\tau}= \sum_{k'} \dfrac{1}{\tau(k\rightarrow k')} = \dfrac{2\pi}{\hbar} \sum_{k'} \langle|V_{kk'}|^2\rangle \, \delta(E_k - E_{k'})
\,
\end{equation}
for surface states of a disordered 3DTI with semi-infinite boundary condition in $\hat{z}$-direction, i.e., $H \longrightarrow \infty$.
Based on Eq. \eqref{energy}, we have
\begin{equation}
\delta(E^+_k - E^+_{k'}) =\dfrac{1}{A_2 \sqrt{1 - \xi^2}}\; \; \delta \left(k'_\perp - \dfrac{E^+ - C - \xi M_0}{A_2 \sqrt{1 - \xi^2}}\right).
\end{equation}
Hence, the resulting total ensemble-averaged mean free time of surface states on the $\hat{z}$-surface reads
\begin{equation} \label{mftz}
\dfrac{1}{\tau} = \dfrac{2\gamma}{\hbar}\dfrac{E^+ - C - \xi M_0}{A_2^2(1-\xi^2)} \dfrac{\alpha}{\beta^2} \, .
\end{equation}
Similarly, for an energy dispersion, to the lowest order of $k$, for  $\hat{y}$-plane surface states,
\begin{equation} \label{energy2}
E^\pm = C + \eta M_0 \pm \sqrt{1 - \eta^2}\;\sqrt{A_2^2 k_x^2 + A_1^2 k_z^2}
 \, ,
\end{equation}
we obtain the total ensemble-averaged inverse mean free time
\begin{equation} \label{mfty}
\dfrac{1}{\tau} \simeq \dfrac{2\gamma}{\hbar}\dfrac{E^+ - C - \eta M_0}{A_2^2(1-\eta^2)} \dfrac{\alpha'}{\beta'^2} \, ,
\end{equation}
where we approximate the Fermi velocity, $v_F = v_{F,x} (\hat{\bf y})$, at this surface based on Eq.~\eqref{energy2} since $A_2 > A_1$.
Note that $\alpha (\beta)$ and $\alpha' (\beta')$ are different values since the depth of the surface states into the bulk in different surfaces are not the same according to the parameters of the Hamiltonian.

According to our mean free time and Fermi velocities derivations, Eqs.~\eqref{fraction} and \eqref{fraction2} yields
\begin{eqnarray}
\left( \frac{\langle S_z \rangle} {d \langle n \rangle /dx}\right)_{- \hat{\bf y}}
=& \left(\dfrac{\eta(A_2 \sqrt{1-\eta^2})^3}{E^+ - C - \eta M_0} \dfrac{\beta'^2}{4\alpha' \gamma}\right)_{- \hat{\bf y}}, \label{eq:Sz} \\
\left( \frac{\langle S_y \rangle}  {d \langle n \rangle / dx} \right)_{+\hat{\bf z}}
=& \left( \dfrac{(A_2 \sqrt{1-\xi^2})^3}{E^+ - C - \xi M_0} \dfrac{\beta^2}{4\alpha \gamma}\right)_{+\hat{\bf z}}. \label{eq:Sy}
\end{eqnarray}

\section{$\eta=0$ case}
\label{app_zero_eta}

\begin{figure}[h!]
	\centerline{\includegraphics[width=0.98\linewidth]{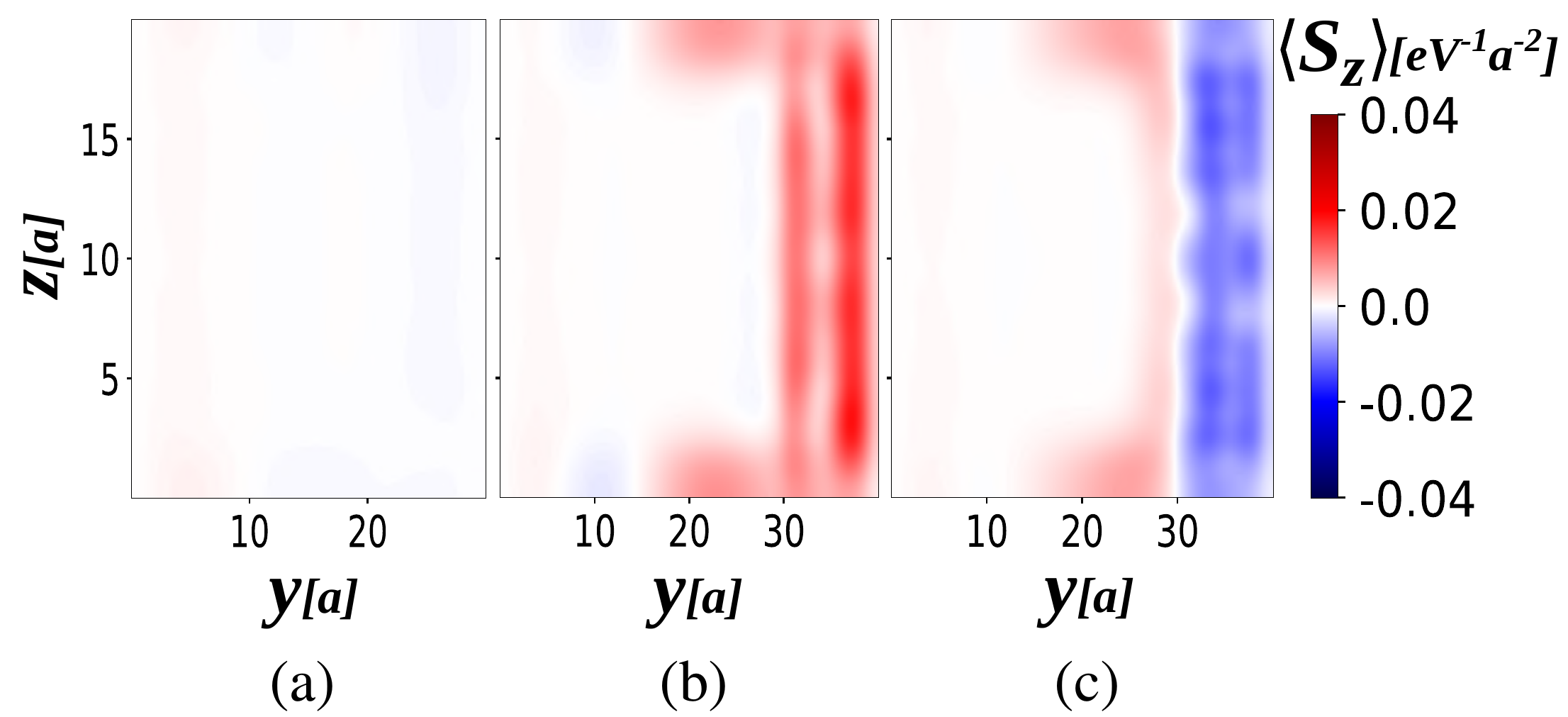}}
	\caption{\label{fig:zeroeta}
		Current-induced spin polarization into a side pocket at the side surface when $\eta=0$. Spatial profile of the averaged spin polarization $\langle S_{z}(y,z) \rangle$ (averaged over 1000 disorder configurations)
		along cross sections in $\hat{\bf x}$ direction. (a) $\langle S_{z}(y,z) \rangle$ corresponds to the system shown in Fig.~\ref{fig:3DTI}(a).
		(b), (c) $\langle S_{z}(y,z) \rangle$ corresponds to the system shown in Fig.~\ref{fig:sideSP}(a).
		In panels (b) and (c) the side pockets, $W_{SP}=10 \, a$, are doped to hole bands ($V_{\text{gate}}=-0.7 \, \rm{eV}$)
		and electron bands ($V_{\text{gate}}=0.7 \, \rm{eV}$), respectively. Common parameters are: $L = 30 \, a$, $W = 30 \, a$, $H = 20 \, a$, $U_{0}=0.5 \, \rm{eV}$, $E_{F} = 0.15 \, \rm{eV}$ which is in the bulk gap. We set $D_2=0$ in all parts of the system.}
\end{figure}

Here we provide figures for the case $D_2 = 0$ leading to $\eta =0$. It is clearly seen that while there is negligible spin accumulation on the side of a 3DTI (Fig.~\ref{fig:zeroeta}(a)), spin extraction is nonnegligible in the side pocket and spin polarization can be switched via a gate potential (see Figs.~\ref{fig:zeroeta}(b) and~\ref{fig:zeroeta}(c)).

\end{document}